\documentclass[floatfix,prb,aps,twocolumn,showpacs,preprintnumbers,amsmath,amssymb]{revtex4}  
\usepackage{graphicx} 
\usepackage{dcolumn} 
\usepackage{exscale} 
\usepackage{color}
\usepackage{bbm}  
\usepackage{amsfonts} 
 
\def\shiftdown#1{#1\llap{\lower.04ex\hbox{#1}}}

\newcommand{\I}{ {\rm Im}} 
\newcommand{\R}{ {\rm Re}} 

\begin{document} 
\title{
  Scalar Wave Propagation in Random, Amplifying Media: Influence of
  Localization Effects on Length and Time Scales and Threshold Behavior} 

\author{Regine Frank$^{1}$ and Andreas Lubatsch$^{2}$}
\affiliation{$^1$ Institut f\"ur Theoretische Festk\"orperphysik, Karlsruhe
  Institute of Technology (KIT), Wolfgang - Gaede - Strasse 1, 76131
  Karlsruhe, Germany\\
$^2$Physikalisches Institut, Universit\"at Bonn, Nussallee 12, 
                 53115 Bonn, Germany}


\begin{abstract} 
\noindent 
We present a detailed discussion of scalar wave propagation and light intensity transport in three dimensional 
random dielectric media with optical gain. The intrinsic length and time scales of such amplifying 
systems are studied and comprehensively discussed as well as the threshold characteristics of
single- and two-particle propagators.
Our semi-analytical theory is based on a self-consistent 
Cooperon resummation, representing the repeated self-interference, and incorporates as well  optical gain
and absorption, modeled in a semi-analytical way by a finite imaginary part of the dielectric function.
Energy conservation  in terms of a generalized Ward identity is taken into account.
\end{abstract} 

\maketitle

\section{Introduction}
Built on a wide story of success and plenty of achievements in science, light propagation and intensity transport 
in multiple-light-scattering random media further provides a lot of remarkable and fascinating features
in theory \cite{Tiggelen0602,Tiggelen0603, sergei, Frank06}
and experiment \cite{maret0601,maret0602}. Special ongoing interest is focused
on optical gain material regarding disordered
as well as periodic structures  \cite{Cao03b,Lagendijk_mie,cao06,cao05,cao04}.

In theoretical physics a profound understanding of this subject requires
the discussion of strong or Anderson localization  \cite{And58} of light, 
a microscopic transport theory in the diffusive limit usually based on conservation laws, 
incorporation of gain or absorption into generalized conservation laws, as well as
observing the occurrence of  a threshold behavior of the light intensity.

Anderson localization
has been shown \cite{gangof4,Vol80} to originate from  repeated self-interference  
of diffusive modes. 
In experiments\cite{Wie97,Sch99,Wie99} the authors have tested this.
Since intensity diffusion  is solely based on energy conservation,
a rigorous and consistent framework is needed to describe the interplay 
between  coherent amplification and localization.
Especially since coherent amplification is expected to enhance 
transmission whereas localization tends to stop light transport altogether.
This interesting subject\cite{Beenakker96} is also discussed in the context of 
random lasing \cite{Cao03b,Cao00}, where it has been
shown by measuring the photon statistics \cite{Cao01STAT}, that the laser emission
is due to coherent feedback and occurs from spatially confined spots  
in the sample. 
A theoretical attempt\cite{Apalkov02} to independently explain such phenomena
has been proposed based on scatterers statistically forming ring resonators within the sample, 
which is able to provide a feedback mechanism.

Despite the issue of transport and localization, also the  onset of lasing behavior,
the crossing of the so-called laser threshold, has  triggered still ongoing 
publications \cite{Nussenzveig,Lai,Vahala,Lagendijk_mie} for the last decades.
On the theoretical side, the difficulty lies in the use of static gain below  threshold,
which is perfectly valid unless the threshold is reached. Beyond the laser threshold,
the laser dynamics have to be explicitely taken into account.
Such a threshold behavior has to be carefully respected and incorporated in a consistent way into any
theory of transport and localization.

In the present paper we clarify this subject by presenting a 
semi-analytical general theory of light propagation, 
localization effects in the presence of optical gain,
and consistently discuss the occurrence and influence of a growth threshold 
on the single particle propagator as well as on the energy density correlation function, 
related to the intensity in the sample.
In this way we set the last stage within the range of linear response theory,
which then will serve as a basis to involve the actual lasing dynamics.
This dynamical behavior is, however, not subject of this article and will
be presented in forthcoming publication.

\section{Model and Theory}

\subsection{Basic Setup}

Systems of significant  experimental relevance \cite{maret0601,maret0602,Cao03b,Lagendijk_mie,cao06,cao05,cao04} 
consist of (almost) spherical scatterers
embedded into a background medium forming some emulsion. 
For a theoretical description we therefore consider identical spherical scatterers located at random positions. 
The scatterers as well as 
the background medium are respectively assumed to be homogeneous and hence will be described by dielectric constants
$\epsilon_s$ and $\epsilon_b$, respectively. Within a  semiclassical context linear absorption and optical gain
will be represented by a finite positive or negative part of the dielectric function, so in general
${\rm Im } \epsilon_s \ne 0 \ne {\rm Im} \epsilon_b$ is assumed.
Throughout the paper we  neglect polarization effects and therefore consider
the {\em scalar} wave equation which has been Fourier transformed from time $t$ to frequency $\omega$ and reads
\begin{equation} 
\label{eq:field} 
\frac{\omega^2}{c^2} \, \epsilon (\vec{r}\,) \Psi_\omega(\vec{r}\,) 
+ \nabla ^2 \Psi_\omega( \vec{r}\,) 
= -i \omega \frac{4\pi}{c^2}  j_\omega(\vec{r}\,)\ , 
\end{equation} 
where $c$ denotes the vacuum speed of light  
and  $j_\omega (\vec{r}\,)$ the current. 
The dielectric constant  
$ \epsilon(\vec{r}\,) = \epsilon_b + \Delta\epsilon\, V(\vec{r}\,)$,  where 
the dielectric contrast has been defined as
$\Delta\epsilon = \epsilon_s - \epsilon_b$, 
describes the arrangement of scatterers through the function 
$V(\vec{r}\,) = \sum_{\vec{R}} S_{\vec{R}}\,(\vec{r}-\vec{R}\,)$, with 
$S_{\vec{R}}\,(\vec{r}\,)$ a localized shape function 
at random locations $\vec{R}$. 
The intensity is then related to the field-field-correlation function
$\langle \Psi(\vec{r},\,t\,) \Psi^*(\vec{r}\,',t\,'\,)\rangle$  here  angular brackets $\langle\ldots\rangle$ denote
ensemble or disorder  average. To calculate the  field-field-correlation the Green's
function formalism is best suited, the  (single-particle) Green's function is 
related to the (scalar) electrical field by
\begin{equation} 
\label{SP_Green_function_field} 
 \Psi(\vec{r},\,t\,) = 
\left\lmoustache  \!  {\rm d}^3r\,'    \right.\!\!
\left\lmoustache  \!  {\rm d}t\,'    \right.
G( \vec{r} \, ,\vec{r} \,'\, ;\,t\,,t'\,)      j(\vec{r}\,'\,,t'\,)\,.
\end{equation}
The Fourier transform  of the retarded, disorder averaged single-particle 
Green's function of Eq.\ (\ref{eq:field}) reads,
\begin{equation}
\label{SP_Green_function}
G_{\vec{q}}^{\omega} = \frac{1}
{\epsilon_b (\omega /c)^2 - \vert \vec{q} \vert^2 - \Sigma^{\omega}_{\vec{q}}} \ ,
\end{equation} 
where the retarded self-energy $\Sigma _{\vec{q}} ^{\omega}$  
arises from scattering off the random ``potential'' 
$-(\omega /c)^2(\epsilon_s - \epsilon_b )V(\vec{r}\,)$.
Using the Green's function the mode density $N(\omega)$ may be expressed as 
$N(\omega)=-(\omega/\pi)\I G_0^{\omega}$, with the abbreviation used throughout 
this publication
$G_0^{\omega}\equiv \int d^3q/(2\pi)^3\, G_{\vec{q}}^{\omega}$.

In order to study the transport of the above introduced field-field-correlation
we consider the so-called 4-point correlation function, 
defined in terms of the non-averaged Green's functions 
$\hat G$, $\hat G^*$ in momentum and frequency space  as 
$\Phi^{\omega}_{\vec{q}\vec{q}^{\prime}\,\,}(\vec{Q},\Omega)= 
\langle \hat G^{\omega_+}_{{\vec{q}}_+{\vec{q}}_+^{\,\prime}} 
        \hat G^{\omega_-\, *}_{{\vec{q}}_-^{\,\prime} {\vec{q}}_-}  
\rangle$.
Here we have introduced the usual \cite{Lubatsch05}  
center-of-mass ($\vec{q}$, $\omega$) and relative  ($\vec{Q}$, $\Omega$) 
frequencies and momenta: 
The variables $\Omega$, $\vec Q$ are associated with the time and 
position dependence of the averaged energy density, with
$\hat Q =\vec Q/|\vec Q|$, while  
$\omega_{\pm} = \omega \pm \Omega /2$ and  
$\vec{q}_{\pm} = \vec{q} \pm \vec{Q}/2$ etc. are the frequencies  
and momenta of in- and out-going waves, respectively. 

The intensity correlation, or disorder averaged 
particle-hole Green's function,
$\Phi^{\omega}_{\vec{q}\vec{q}^{\prime}}(\vec{Q},\Omega)$
obeys the so-called 
Bethe-Salpeter equation 
\begin{eqnarray}
\label{bethe_eq}
\!\!\!\!\!\!\!\!\!\!\!\!\!\!\!\!\!\!\!\!\!\!\!\!\!\!\!\!\!\!\!\!\!\!\!\!\Phi_{\vec{q}\vec{q}'}\, 
=\,
G^R_{q_+}(\omega_+)G^A_{q_-}(\omega_-) {\color{white}0000000000000000}\nonumber\\
\!
\qquad{\color{white}00000000}\left[ 
\delta(\vec q - \vec q\prime)  
+ 
\left\lmoustache     \frac { {\rm d}^3q\,''\,}    { (2\pi)^3 } \right. 
\gamma_{q\,q\,''\,}\Phi_{\vec{q}\,''\,\vec{q}\,'\,}
\right]\,.
\end{eqnarray}
By utilizing the known averaged single particle Green's function, {\em c.f.} Eq. (\ref{SP_Green_function}), on the left-hand side
of Eq. (\ref{bethe_eq}) the Bethe-Salpeter equation may be rewritten as  kinetic equation, see, e.g., 
Ref.\ \cite{Lubatsch05}, 
\begin{eqnarray} 
\label{boltzmann} 
\left[ 
\omega\Omega\frac{\R{\epsilon_b}}{c^2} 
- Q\, (\vec{q}\cdot\hat{Q}) +\frac{i}{c^2\tau ^2} 
\right] 
\Phi^{\omega}_{\vec{q}\vec{q}^{\prime}} 
&=& \nonumber\\ 
&&\hspace*{-4.7cm}- i \I  G^{\omega}_{\vec{q}} 
\left[ \delta(\vec q -\vec q\prime) + 
\left\lmoustache  \frac{{\rm d}^3 q^{\prime\prime}}{(2 \pi)^3} \right. 
 \gamma^{\omega}_{\vec{q}{\vec{q}^{\prime\prime}}} 
\Phi^{\omega}_{\vec{q}^{\prime\prime}\vec{q}^{\prime}} 
\right]. 
\end{eqnarray} 
In order to analyze the correlation function's long-time 
($\Omega \to 0$) and long-distance ($ \vert \vec{Q} \vert \to 0$)  
behavior, terms of $O(\Omega^2, Q^3, \Omega Q)$ have been  
neglected here and throughout this paper.  
Eq.\ (\ref{boltzmann}) contains both, the {\em total}  
quadratic momentum relaxation rate $1/\tau^{2}=c^2\,\I ( \epsilon_b  
\omega^2/c^2-\Sigma ^{\omega})$ (due to absorption/gain in the  
background medium as well as impurity scattering) and the 
irreducible two-particle vertex function 
$\gamma^{\omega}_{\vec{q} \vec{q}^{\,\prime}}(\vec{Q},\Omega)$.

To solve this equation, the technique of expansion into moments  is used.
The technical details of this expansion are discussed in the following
subsection B. The Reader not interested in such details may skip this and
readily proceed to subsection C.

Furthermore it is to be noted that 
the energy conservation is implemented into the solution 
of the Bethe-Salpeter equation in a field theoretical sense  
by a Ward identity (WI) which has been derived for the photonic case in 
Ref.\ \onlinecite{Lubatsch05}, and which for scalar waves takes the exact 
form 
\begin{eqnarray} 
\label{Ward} 
\Sigma^{\omega_+}_{\vec{q}_+} - \Sigma^{\omega_-\, *}_{\vec{q}_-} 
 \!\!&-& \!\! 
\left\lmoustache  \!\!\frac {{\rm d}^3 q^{\prime}}{(2\pi)^3} \right. 
\left[G^{\omega_+}_{{\vec{q}}_+^{\,\prime}} - G^{\omega_-\, 
  *}_{{\vec{q}}_-^{\,\prime}} \right] \,  
{ \gamma}^{\omega}_{{\vec{q}}^{\,\prime}{{\vec{q}}}}({\vec{Q}},\Omega) 
\\ 
 \! \!\!&=& \!\!\! 
f_{\omega}(\Omega)\! 
\left[\! 
\R{\Sigma}^{\omega}_{{\vec{q}}} 
 \!+\!\! 
\left\lmoustache  \!\!\frac {\rm d^3 q^{\prime}}{(2\pi)^3} \right. 
\R {G}^{\omega}_{{\vec{q}}^{\,\prime}} \,  
{ \gamma}^{\omega}_{{\vec{q}}^{\,\prime}{{\vec{q}}}}({\vec{Q}},\Omega) 
\right]\! . 
\nonumber 
\end{eqnarray} 
The right-hand side of Eq.\ (\ref{Ward}) represents reactive effects 
(real parts), originating from the explicit $\omega^2$-dependence of the 
photonic random ``potential''. In conserving 
media ($\I \epsilon _b = \I \epsilon _s  =0$) these terms renormalize the 
energy transport velocity $v_{\mbox{\tiny E}}$ relative to the 
average phase velocity  $c_p$  without 
destroying the diffusive long-time behavior.\cite{Kro93,Lubatsch05} 
In presence of loss 
or gain, however, these effects are enhanced via the prefactor  
$f_{\omega}(\Omega)= 
(\omega\Omega \R \Delta\epsilon + i\omega^2 \I \Delta\epsilon )/ 
(\omega^2     \R \Delta\epsilon + i\omega\Omega \I \Delta\epsilon )$, 
which now does not vanish in the limit 
$\Omega \to 0$.

\begin{figure}[t] 
\begin{center} 
\includegraphics[width=0.75\linewidth]{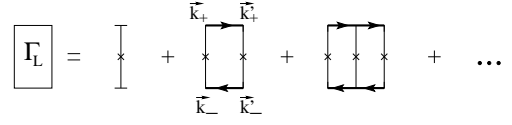} 
\end{center} 
\vspace*{-0.4cm}
\caption{
         Ladder approximation of the total particle-hole vertex.
	 The diagrams on the left-hand side form a geometrical series
	 and may therefore easily be summed up analytically.
         }
\label{particle_hole_ladder_pic}
\end{figure}

\subsection{Expansion of Two-particle Green's Function into Moments}

In order to extract a diffusion pole  structure out of the Bethe-Salpeter equation Eq. (\ref{bethe_eq}),
the  correlator or equivalently  the $\vec{q}\,'\,$ integrated correlator
\begin{eqnarray}
\label{Phi_q}
\Phi_{\vec{q}}
=
\left\lmoustache   \! \!  \frac { {\rm d}^3q\,'\,}    { (2\pi)^3 } \right.\! \!  
\Phi_{\vec{q\,}\vec{q}\,'\,} 
\end{eqnarray}
has to be decoupled from the momentum dependent prefactors with the help of some approximation scheme.
In this subsection we discuss this procedure in analogy to the argumentation for electronic correlations
presented in reference \cite{Kopp}.
Such an approximation must obey the results of the so-called ladder approximation as well as it must incorporate
the set of physical relevant variables involved in  observed phenomena.
In a first step we use the bare first two moments of the correlation function $\Phi_{\vec{q}}$
defined according to
\begin{eqnarray}
\label{density_density}
\Phi_{\rho\rho}(\vec{Q},\Omega)
&=&
\left\lmoustache   \!  \frac { {\rm d}^3q}    { (2\pi)^3 } \right.
\!\!
\left\lmoustache   \!  \frac { {\rm d}^3q\,'} { (2\pi)^3 } \right.
\Phi_{\vec{q}\,\vec{q}\,'\,}
\\
\label{density_current}
\Phi_{j\rho}(\vec{Q},\Omega)
&=&
\left\lmoustache   \! \frac { {\rm d}^3q}  { (2\pi)^3 } \right.
\!\!
\left\lmoustache   \!\frac { {\rm d}^3q\,'}{ (2\pi)^3 } \right.
(\vec{q}\cdot\hat{Q})
\Phi_{\vec{q}\,\vec{q}\,'\,},
\end{eqnarray}
respectively. 
The second step is to recognize that
these bare moments are related  to  physical quantities,
the energy density correlation $P^{\omega}_{\mbox{\tiny E}}(\vec{Q},\Omega)$
and the current-density-correlation $J^{\omega}_{\mbox{\tiny E}}(\vec{Q},\Omega)$,
by dimensional prefactors:
\begin{eqnarray}
\label{density_density_physical}
P^{\omega}_{\mbox{\tiny E}}(\! \vec{Q},\Omega )\!\! &=&\! \! 
\left[\!
\frac{\omega}{c_{\mbox{\tiny p}}}
\!\right]^2\!\!\!  \Phi_{\rho\rho}
\,\,\,\,
\Leftrightarrow
\Phi_{\rho\rho}
\!=\! \left[\!
\frac{c_{\mbox{\tiny p}}}{\omega}
\!\right]^2\!\!P^{\omega}_{\mbox{\tiny E}}(\! \vec{Q},\Omega )
\\
\label{density_cuurent_physical}
J^{\omega}_{\mbox{\tiny E}}(\! \vec{Q},\Omega )\!\! &=& \!\! 
\left[\!
\frac{\omega v_{\mbox{\tiny E} }} {c_{\mbox{\tiny p}}}
\!\right]\!
 \Phi_{j\rho}
\Leftrightarrow
\Phi_ {j\rho}
\! =\! 
\left[\! 
\frac {c_{\mbox{\tiny p}}} {\omega v_{\mbox{\tiny E} }} 
\! \right]\!
J^{\omega}_{\mbox{\tiny E}}(\! \vec{Q},\Omega )\, .
\end{eqnarray}
The projection of the correlator $\Phi_{\vec{q}}$, Eq. (\ref{Phi_q}), onto the bare 
moments $\Phi_ {\rho\rho}(\vec{q},\Omega)$ as defined in Eq. (\ref{density_density}),
and  $\Phi_ {j\rho}(\vec{q},\Omega)$, shown in  Eq. (\ref{density_current}), is therefore given by
\begin{eqnarray}
\label{intensity_projection}
\left\lmoustache   \! \frac { {\rm d}^3q\,'}{ (2\pi)^3 } \Phi_{\vec{q}\vec{q}\,'}  \right.
&\!\!=\!\!&
\frac{A(\vec{q}\,)}
     {
       \! \left\lmoustache   \! \!  \frac { {\rm d}^3q'}{ (2\pi)^3 }  \right. \! \! A(\vec{q}\,')}
\Phi_ {\rho\rho}(\vec{Q},\Omega)
\\
&\!\!+\!\!&
\frac{B(\vec{q}\,)  (\vec{q}\cdot\hat{Q}) }
     {
        \! \left\lmoustache   \! \!  \frac { {\rm d}^3q'}{ (2\pi)^3 } \right. \! \! 
	B(\vec{q}\,')(\vec{q}\,'\cdot\hat{Q})^2}
\Phi_ {j\rho}(\vec{Q},\Omega),\nonumber
\end{eqnarray}
where the projection coefficients $A(\vec{q}\,)$ and $B(\vec{q}\,)$  are to be determined in the following.
For obvious reasons in this expansion the bare moments may be substituted by their physical counterparts
energy density $P^{\omega}_{\mbox{\tiny E}}$ in Eq. (\ref{density_density_physical}) and 
current density $J^{\omega}_{\mbox{\tiny E}}(\vec{Q},\Omega)$ from Eq. (\ref{density_cuurent_physical}).
The expansion coefficients $A(\vec{q}\,)$ and $B(\vec{q}\,)$ in Eq. (\ref{intensity_projection})
behave uncritically under localization, so they can be determined
using the simple ladder approximation, where all expressions are known exactly.
The ladder approximation of the two-particle vertex function is explicitely illustrated in Fig. \ref{particle_hole_ladder_pic}.
In the following we use this approximation and demonstrate how to obtain the expansion coefficients from it.
In ladder approximation the  zeroth bare moment is given by:
\begin{eqnarray}
\Phi_ {\rho\rho}^L (\vec{Q},\Omega)
&=&
\left\lmoustache   \!  \frac { {\rm d}^3q}{ (2\pi)^3 }   \right.\! \! 
\left[ 
G_{\vec{q}_+}(\vec{Q},\Omega)G^*_{\vec{q}_-}(\vec{Q},\Omega)
\right]
^2 \Gamma_L\nonumber\\
\label{density_ladder}
&=&
\frac 1{\tilde\gamma_0^2}\Gamma_L,
\end{eqnarray}
the superscript $L$ refers to the ladder approximation and  
in the last step the product 
$
\left\lmoustache   \!  
\frac { {\rm d}^3q}
           { (2\pi)^3 } \right.\! \! 
\left[ 
G_{\vec{q}_+}(\vec{Q},\Omega)G^*_{\vec{q}_-}(\vec{Q},\Omega)
\right]
^2 
$
has been expanded up to linear order in $\vec{q}$ 
furthermore the renormalized vertex $\tilde\gamma_0$ is given by
\begin{eqnarray}
\tilde\gamma_0
&\!\!=\!\!&
\gamma_0
\!+\!
f_{\omega}(\Omega)
\frac
{
\left(\!
{\rm Re\,}\gamma_0  G_0
\!+\!
{\rm Re\,} \Sigma \!
\right)
}
{
{\rm Im\,} G_0
}
\!- \!
\frac
{
\omega^2 {\rm Im\,} \epsilon_b
}
{
 {\rm Im\,} G_0
}
\end{eqnarray}
where $\gamma_0$ is the bare vertex and 
$f_{\omega}(\Omega)$ arising from the Ward identity
has been defined in Eq. (\ref{Ward}).
Within the simple ladder approximation the bare  moment $\Phi^L_{j\rho}(\vec{Q},\Omega)$
defined in Eq. (\ref{density_current}) is thus given by 
\begin{eqnarray}
\Phi^L_{j\rho}(\!\vec{Q},\Omega)
\!\!=\!\!
\! \left\lmoustache   \!  \frac { {\rm d}^3q}   { (2\pi)^3 }\right.\! \! 
(\vec{q}\cdot\hat{Q})
G_{\vec{q}_+}\!G^*_{\vec{q}_-}
\! \! \left\lmoustache   \!   \frac { {\rm d}^3q\,'\,} { (2\pi)^3 } \right.\! \! 
G_{\vec{q}\,'_+}\!G^*_{\vec{q}\,'_-}\!\!
\Gamma_L\,.
\end{eqnarray}
Following the above strategy and expanding the product $G_{\vec{q}\,'_+}G^*_{\vec{q}\,'_-}$
under the second integral up to first order in $\vec{q\,}'$
one obtains the expression
\begin{eqnarray}
\Phi^L_ {j\rho}(\vec{Q},\Omega)
=
\frac 1{\tilde\gamma_0}
\Gamma_L
\! \left\lmoustache   \!  \frac { {\rm d}^3q} { (2\pi)^3 }\right.\! \! 
(\vec{q}\cdot\hat{Q})
G_{\vec{q}_+}G^*_{\vec{q}_-}
\end{eqnarray}
By now employing the same expansion
to the remaining product of the Green's function one eventually finds
\begin{eqnarray}
\label{current_ladder}
\Phi^L_ {j\rho}(\vec{Q},\Omega)
=
\frac {\Gamma_L} {\tilde\gamma_0}
\! \left\lmoustache   \!  \frac { {\rm d}^3q}  { (2\pi)^3 }\right.\! \! 
(\vec{q}\cdot\hat{Q})
\frac 12
\frac 
{\Delta G_{\vec{q}}^2  (\vec{q}\cdot\hat{Q}) Q  }
{\tilde\gamma_0 \Delta G_0},
\end{eqnarray}
where the abbreviation $\Delta G \equiv G - G^*$ has been introduced and will be used 
throughout this paper.

In the next step of determining the expansion coefficients $A(\vec{q}\,)$ 
and $B(\vec{q}\,)$ defined in Eq. (\ref{intensity_projection}) we go back to the 
field-field correlation function
$ 
\Phi_{\vec{q}\vec{q}\,'\,}$.
Within the uncritical ladder approximation the two particle Green's function
is given by
\begin{eqnarray}
\label{some_phi_ladder}
\! \left\lmoustache   \!  \frac { {\rm d}^3q\,'}{ (2\pi)^3 } \right.\! \! 
\Phi_{\vec{q}\vec{q}\,'}
=
\left[
G_{\vec{q}_+}G^*_{\vec{q}_-}
\right]
\Gamma_L
\! \left\lmoustache   \!  \frac { {\rm d}^3q\,'}  { (2\pi)^3 }\right.\! \! 
G_{\vec{q}\,'_+}G^*_{\vec{q}\,'_-}.
\end{eqnarray}
Employing again the momentum expansion of the single-particle Green's function
the above equation, Eq. (\ref{some_phi_ladder}) can be simplified to yield
\begin{eqnarray}
\label{intensity_ladder}
\Phi_{\vec{q}}
=
\frac
{\Delta G_{\vec{q}}}
{\tilde\gamma_0^2 \Delta G_0}\Gamma_L
+
\frac 12
\frac
{\Delta G_{\vec{q}}^2 (\vec{q}\cdot\hat{Q}) Q}
{\tilde\gamma_0^2 \Delta G_0}\Gamma_L.
\end{eqnarray}
Finally we are in the position to start putting things together.
By using the above given momentum expansion,
Eq. (\ref{intensity_ladder}), together with the expressions given 
in Eq. (\ref{current_ladder}) and in Eq. (\ref{density_ladder})
in conjunction with the proposed projection, or expansion into moments,
Eq. (\ref{intensity_projection}),  the following relation is eventually obtained
\begin{eqnarray}
\label{comp_of_coeff}
&&
\frac
{\Delta G_{\vec{q}}}
{\tilde\gamma_0^2 \Delta G_0}\Gamma_L
+
\frac 12
\frac
{\Delta G_{\vec{q}}^2 (\vec{q}\cdot\hat{Q}) Q}
{\tilde\gamma_0^2 \Delta G_0}\Gamma_L
\\\nonumber
&&=\!
\frac{A(\vec{q}\,)}
     {
       \! \left\lmoustache   \! \! \frac { {\rm d}^3q'}  { (2\pi)^3 }  \right.\! \!   A(\vec{q}\,')}
\frac 1{\tilde\gamma_0^2}\Gamma_L
\\
&&+
\frac{B(\vec{q}\,)  (\vec{q}\cdot\hat{Q}) }
     {
       \! \left\lmoustache   \! \!\frac { {\rm d}^3q'}{ (2\pi)^3 }  \right.\! \!  
       B(\vec{q}\,')(\vec{q}\,'\cdot\hat{Q})^2}
\frac {\Gamma_L}{\tilde\gamma_0}
 \! \left\lmoustache   \! \frac { {\rm d}^3q} { (2\pi)^3 } \right.\! \!  
(\vec{q}\cdot\hat{Q})
\frac 12
\frac 
{\Delta G_{\vec{q}}^2  (\vec{q}\cdot\hat{Q}) Q  }
{\tilde\gamma_0 \Delta G_0}\,.
\nonumber
\end{eqnarray}
By comparison of coefficients in the above relation, Eq. (\ref{comp_of_coeff}),
the  demanded coefficients $A({\vec{q}}\,)$ and $B({\vec{q}}\,)$ of the 
expansion into moments, Eq. (\ref{intensity_projection}), can now be determined 
to be
\begin{eqnarray}
A({\vec{q}}\,)
=\Delta G_{\vec{q}}
\qquad
B({\vec{q}}\,)
=\Delta G_{\vec{q}}^2.
\end{eqnarray}
Employing those expressions for the expansion coefficients, one may eventually express
the two-particle correlator $\Phi_{\vec{q}\vec{q}\,'}$ in the following way
\begin{eqnarray}
\label{decoupled_projection_physical}
\! \left\lmoustache   \!  \frac { {\rm d}^3q'}{ (2\pi)^3 }  \right.\! \!  
\Phi_{\vec{q}\vec{q}\,'}
&=&
\frac{
         \Delta G_{\vec{q}}
     }
     {
       \left(
       \frac{\omega}{c_{\mbox{\tiny p}}}
       \right)^2 
       \! \left\lmoustache   \! \! 
       \frac { {\rm d}^3q\,'} { (2\pi)^3 } \right.\! \!  
       \Delta G_{\vec{q}\,'}   
     }
P^{\omega}_{\mbox{\tiny E}}(\vec{Q},\Omega)
\\
&&+
\frac {
         \Delta G_{\vec{q}}^2 (\vec{q}\cdot\hat{Q}) 
      }
      {
	\left(
	\frac{\omega v_{\mbox{\tiny E} }} {c_{\mbox{\tiny p}}}
	\right)
	\! \left\lmoustache   \! \! 
	\frac { {\rm d}^3q\,'}{ (2\pi)^3 } \right.\! \!  
	\Delta G_{\vec{q}\,'}^2 (\vec{q}\,'\cdot\hat{Q})^2
      }
J^{\omega}_{\mbox{\tiny E}}(\vec{Q},\Omega).
\nonumber
\end{eqnarray}
The above expression, Eq. (\ref{decoupled_projection_physical}), 
represents the complete expansion of the intensity correlator into 
its moments.
This will be used in the next subsection 
to decouple and therefore solve the Bethe-Salpeter equation.

\subsection{General Solution of the Bethe-Salpeter Equation}

The disorder averaged intensity correlation, the two-particle Green's function,
obeys the Bethe-Salpeter equation, see Eq. (\ref{bethe_eq})
\begin{eqnarray}
\Phi_{\vec{q}\,\vec{q}\,'\,} 
\!\!&=&\!\! 
G_{q_+}^{\omega_+}G^{*\,\omega_-}_{q_-}
\!
\left[ 
1 \! \!  
+ \! \!  \!  
\left\lmoustache   \! \!  \frac { {\rm d}^3q\,''\,}    { (2\pi)^3 } \right.\! \!  
\gamma_{q\,q\,''\,}\Phi_{\vec{q}\,''\,\vec{q}\,'}
\right]
\end{eqnarray}
as already discussed the Bethe-Salpeter equation may be rewritten
into the kinetic equation  given in Eq. (\ref{boltzmann})
\begin{eqnarray}
\label{kinetic}
&&\!\!\!\!\!\!\!\!\!\!\!\!\!\!\!\!\!\!\!
\left[ \omega\Omega 2 {\rm Re\,}\epsilon- Q \left(\vec{q}\cdot\hat{Q}\right) + \Delta \Sigma -\omega^2 \Delta\epsilon  \right] 
\Phi_{\vec{q}}
\nonumber\\
&&\qquad
=
\Delta G_{\vec{q}} + 
\left\lmoustache   \! \!  \frac { {\rm d}^3q\,'}    { (2\pi)^3 } \right.\! \!  
\Delta G_{\vec{q}} \gamma_{\vec{q}\vec{q}\,'\,}\Phi_{\vec{q}\,'\,}\,.
\end{eqnarray}
To find the solution of Eq. (\ref{kinetic}),
in a first step one 
sums in Eq. (\ref{kinetic}) over momenta $\vec{q}$, incorporates the generalized  
Ward identity as given in Eq. (\ref{Ward}) and subsequently expands the obtained result 
for small internal momenta $Q$ and internal frequencies $\Omega$.
It is also essential to employ the decoupling shown in Eq. (\ref{decoupled_projection_physical}).
Eventually after some algebraic manipulations the generalized continuity equation for the 
energy density is found to be
\begin{eqnarray} 
\Omega P^{\omega}_{\mbox{\tiny E}} + Q J^{\omega}_{\mbox{\tiny E}} =  
\frac{4\pi i \,\omega\, N(\omega )} 
       {g^{(1)}_{\omega}\left[ 1 + \Delta(\omega) \right] c_p^{2}} 
\!\!&+&\!\! \frac{i [g^{(0)}_{\omega} + \Lambda(\omega) ]} 
       {g^{(1)}_{\omega}\left[ 1 + \Delta(\omega) \right] } 
       P^{\omega}_{\mbox{\tiny E}}  
\nonumber\\  
\label{continuityL}
\end{eqnarray}
which represents energy conservation in the presence of optical gain
and/or absorption.

Within the standard solution procedure the next step is to obtain
a linearly independent equation which also relates the energy density
$P^{\omega}_{\mbox{\tiny E}}$ and the current density $J^{\omega}_{\mbox{\tiny E}}$.
This is realized in a similar way to above,
foregoing
one first multiplies the  kinetic equation, Eq.(\ref{kinetic}), 
by the projector $\left[\vec{q}\cdot\hat{Q}\right]$ 
and then follows the above outlined recipe to eventually obtain
the wanted second relation, this is the  so-called current relaxation equation
\begin{eqnarray} 
\label{CDR} 
\left[\omega\Omega\frac{\R{\epsilon_b}}{c^2} 
      +\frac{i}{c^2\tau ^2}+iM(\Omega) 
\right] 
J^{\omega}_{\mbox{\tiny E}} \!\!&+&\!\! 
\tilde A\, Q P_{\mbox{\tiny E}}^{\omega} =0\  
,
\end{eqnarray} 
relating as demanded energy density $P^{\omega}_{\mbox{\tiny E}}$ 
and energy density current $J^{\omega}_{\mbox{\tiny E}}$
and furthermore  introduces the so-called memory function $M(\Omega)$
according to
\begin{eqnarray} 
\label{M_Omega}
M\!(\Omega ) 
\!=\!
\frac
{i\!\!
\left\lmoustache  \!\!\frac {{\rm d}^3 q}{(2\pi)^3} \right.
\!\!\!
\left\lmoustache  \!\!\frac {{\rm d}^3 q\,'}{(2\pi)^3} \right. 
\!\!
[\vec{q}\!\cdot\!\hat{Q}] 
\Delta G_{\vec{q}}^{\omega}
\gamma^{\omega}_{\vec{q}\vec{q}\,'}
(\Delta G_{\vec{q}\,'}^{\omega})^2
[\vec{q}\,'\!\cdot\!\hat{Q}] 
}
{
\left\lmoustache  \!\!\frac {{\rm d}^3 q}{(2\pi)^3} \right. 
[\vec{q}\!\cdot\!\hat{Q}]^2 (\Delta G_{\vec{q}}^{\omega})^2 
}.
\end{eqnarray} 
where $\gamma^{\omega}_{\vec{p}\vec{p}'}\equiv \gamma^{\omega}_{\vec{p}\vec{p}'}(\vec{Q},\Omega)$
is the total irreducible two-particle  vertex, which will be discussed in more detail in the following subsection.

So far, two independent equations, Eq. (\ref{continuityL}) and Eq. (\ref{CDR}),
have been obtained, both of them relating 
the current density $J^{\omega}_{\mbox{\tiny E}}$ and density $P^{\omega}_{\mbox{\tiny E}}$.
Therefore one may now eliminate one of the two variables in this
linear system of equations.
One chooses to  combine the two equation to find an  expression for the energy density 
\begin{eqnarray} 
\label{P_E} 
P_{\mbox{\tiny E}}^{\omega}(Q,\Omega) = 
\frac{4\pi i N(\omega) /  
(g^{(1)}_{\omega}\left[ 1 + \Delta(\omega) \right] c_p^{2})} 
{\Omega + i Q^2 D + i \xi_a^{-2}D}\ , 
\end{eqnarray} 
exhibiting the expected diffusion pole structure for
non-conserving  media, i.e. in the denominator of Eq. (\ref{P_E})
there appears an additional term as compared to the case of conserving media.
This is the term $ \xi_a^{-2}D$, sometimes referred to as the  mass term,  accounting for loss 
(or gain) to the intensity not being due to diffusive relaxation.
In  Eq. (\ref{P_E}) also  the generalized, 
$\Omega$-dependent diffusion coefficient $D(\Omega)$ has been
introduced via the relation 
\begin{eqnarray} 
\label{D_omega_full} 
D(\Omega)  
\left[1 - i \, \Omega \omega \tau ^2 \R\epsilon_b  \right] =   
D_0^{tot} -c^2\tau^2 D(\Omega ) M(\omega ).
\end{eqnarray} 
Furthermore,  Eq. (\ref{P_E}) also introduces
the absorption or gain induced growth or absorption
scale $\xi_a$ of the diffusive modes, 
\begin{eqnarray}  
\label{xi_a} 
\xi_a^{-2}&=& 
\frac
{r_{\epsilon} A_{\epsilon} -2\omega^2\I \epsilon_b} 
{2\R\epsilon_b-A_{\epsilon}B_{\epsilon}/\omega}\ 
\frac {1} {\omega D(\Omega)},
\end{eqnarray} 
which is to be distinguished from the single-particle or amplitude
absorption or amplification length.
The diffusion constant without memory effects in Eq. (\ref{D_omega_full}),  
$D_0^{tot}=D_0+D_b+D_s$, consists of the bare diffusion constant \cite{Kro93}, 
\begin{eqnarray} 
D_0 = \frac  
{2v_{\mbox{\tiny E}} c_p} 
{ \pi N(\omega)} 
\left\lmoustache  \!\!\frac {{\rm d}^3 q}{(2\pi)^3} \right. 
[\vec{q}\cdot\hat{Q}]^2 (\I G_{\vec{q}}^{\omega})^2 
\label{Dbare} 
\end{eqnarray} 
and renormalizations from absorption or gain in the background medium 
($D_b$) and in the scatterers ($D_s$), 
\begin{eqnarray} 
\label{D_B} 
D_b =  \left(\omega\tau \right)^2  \, \I\epsilon_b\, \tilde{D}_0/4 \ ,
\qquad 
D_s = r_{\epsilon}A_{\epsilon}\tau^2 \tilde{D}_0/8 \ , 
\end{eqnarray} 
where $\tilde D_0$ is the same as in Eq.\ (\ref{Dbare}), 
with $(\I G_{\vec{q}}^{\omega})^2$ replaced by 
$\R (G_{\vec{q}}^{\omega\,2})$. In the above Eqs. (\ref{xi_a})-(\ref{D_B}) 
the following short-hand notations have 
been introduced, 
\begin{eqnarray}
u _{\epsilon}&=&\frac {\I (\Delta\epsilon \Sigma^{\omega})}
                 {\I (\Delta\epsilon G_0^{\omega}) }
\ ,\qquad\qquad
r_{\epsilon} = {\I \Delta\epsilon}/{\R \Delta\epsilon},
\nonumber\\
A_{\epsilon} &=& 2 [u _{\epsilon} \R G_o +  \R \Sigma_o] \nonumber\\
B_{\epsilon} &=& \frac{(\R\Delta\epsilon)^2+(\I\Delta\epsilon)^2}
{2\omega^2(\R\Delta\epsilon)^2}.
\nonumber
\end{eqnarray}

\subsection{Vertex Function and Self-consistency}

\begin{figure}[t] 
\begin{center} 
\includegraphics[width=0.75\linewidth]{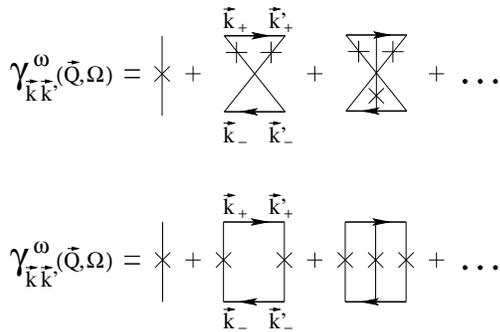} 
\end{center} 
\vspace*{-0.4cm}
\caption{
         The upper panel shows a diagrammatic expansion
	 of the irreducible two-particle vertex $\gamma$. The lower panel 
	 displays the disentangled Cooperon with changed momentum arguments
	 as discussed in the text below.
         }
\label{Cooperon_Flip} 
\end{figure}

From equations Eq. (\ref{M_Omega}) and Eq. (\ref{D_omega_full}) it is clear that
the energy density or two-particle function given in Eq. (\ref{P_E}) still depends
on the full two-particle vertex $\gamma^{\omega}_{{\vec{q}}^{\,\prime}{{\vec{q}}}}$.
Before discussing the vertex function, we want to briefly 
recall some arguments concerning dissipation.
As a simplified  argument to better understand the physical content of
the presented systems, one might consider a damped harmonic oscillator. 
The damping term clearly introduces dissipation as it breaks time reversal symmetry.
However, the time reversed solution is still damped with the very same 
damping constant. 
This shows that the dissipation rate itself is invariant under time reversal, 
which at first sight might sound surprising.

Bearing this in mind one may
carefully analyze the vertex  $\gamma^{\omega}_{{\vec{q}}^{\,\prime}{{\vec{q}}}}$ 
for the self-consistent calculation of 
$M(\Omega )$ \cite{Vol80,Kro90}, exploiting time reversal symmetry of 
propagation in the active medium.
In the long-time limit ($\Omega\to 0$) the dominant contributions to 
$\gamma^{\omega}_{{\vec{q}}^{\,\prime}{{\vec{q}}}}$ are the same 
maximally crossed diagrams (Cooperons) as for conserving media,  
which may also be disentangled. 
In Fig. (\ref{Cooperon_Flip}) the disentangling of the  Cooperon into 
the regular diffusion ladder is demonstrated. The internal momentum argument of the 
disentangled irreducible vertex function in the second line
of  Fig. (\ref{Cooperon_Flip})
is replaced by the new momentum $\vec{Q} = \vec{k} + \vec{k}'$ .
By the described  procedure $\gamma^{\omega}_{{\vec{q}}^{\,\prime}{{\vec{q}}}}$ 
now acquires the absorption (gain)-induced decay (growth) rate 
$\xi_a^{-2}D$. 
Finally the memory kernel  $M(\Omega)$ reads   
\begin{eqnarray} 
M(\Omega ) &=& 
- \frac{(2v_{\tiny E}c_p)^2\ 
u_{\epsilon} \left[ 
2\pi \omega u_{\epsilon} N(\omega ) + r_{\epsilon}A_{\epsilon}  
- 2\omega^2\I\epsilon_b 
\right] 
}
{\pi \omega N(\omega) D_0 D(\Omega)} 
\nonumber\\ 
&& 
\hspace*{-1.5cm} 
\times   
\left\lmoustache \!\frac {{\rm d}^3 q}{(2\pi)^3} \right. 
\left\lmoustache \!\frac {{\rm d}^3 q'}{(2\pi)^3}\right. 
\frac{ 
[\vec{q}\cdot\hat{Q}] 
|\I G_q| \left(\I G_{q'}\right)^2 
[\vec{q}\,'\cdot\hat{Q}] 
} 
{\frac{-i\Omega}{D(\Omega)} + \left(\vec{q}+\vec{q}\,' \right)^2 +  
\xi_a^{-2} }\, . 
\label{MD} 
\end{eqnarray} 
Eqs.\ (\ref{D_omega_full})-(\ref{MD})  
constitute the self-consistency equations for the diffusion coefficient 
$D(\Omega )$ and the growth/decay length $\xi _a$ in presence of absorption or 
gain.

\section{Results and Discussion}

\subsection{Diffusion Constant and its Renormalization}

In Eq. (\ref{D_omega_full}) the diffusion coefficient $D(\Omega, \omega)$
has been shown to consist of a memory induced part already discussed in the last section
and a part $D_0^{tot}(\omega)$, given by the so-called bare diffusion constant $D_0$ defined
in Eq. (\ref{Dbare}) and additional contributions solely due to a finite gain or absorption.
These renormalizations of the diffusion constant are 
\begin{eqnarray} 
\label{D_B_NR} 
D_b =  \left(\omega\tau \right)^2  \, \I\epsilon_b\, \tilde{D}_0/4 \ ,
\qquad 
D_s = r_{\epsilon}A_{\epsilon}\tau^2 \tilde{D}_0/8 \ .
\end{eqnarray} 
Since we are interested in amplifying systems with negligible absorption
we now  want to discuss setups with a finite optical gain coefficient inside 
the scatterers only, which are themselves embedded in a conserving media e.g. air. 
In the background ${\rm Im \,} \epsilon_b$  is identical zero and therefore also $D_b $. 
The described  systems are of strong experimental interest \cite{Cao00,Lagendijk_mie} and are still not
completely understood \cite{Cao03b}.

Before  starting with various  examples we want to point out that in dealing
with optically amplifying media one has to choose
parameters  carefully, guaranteeing that the system remains below its laser threshold.
This will be discussed in detail in a following subsection.
The presented numerical results utilize parameter sets 
which do show below-threshold behavior within the considered frequency range.
In particular we present results for three characteristic parameter sets,
the setup is an optically neutral background medium like air ($\epsilon_b = 1.0$),
spherical scatterers (filling fraction $\nu = 30\%$) with three different 
gain strengths ($\epsilon_{scat} = 10.0 - \{0, 1e-4, 1e-2\} I $).
The system with purely real dielectric functions, i.e. conserving media, serves as a reference system,
and the gain is either  typical (${\rm Im \,}  \epsilon_{scat} = -1e-4$) or rather
large  in magnitude, here ${\rm Im \,}  \epsilon_{scat} = -1e-2$.

\begin{figure}[t] 
\begin{center} 
\vspace*{1.5em}
\includegraphics[width=0.90\linewidth]{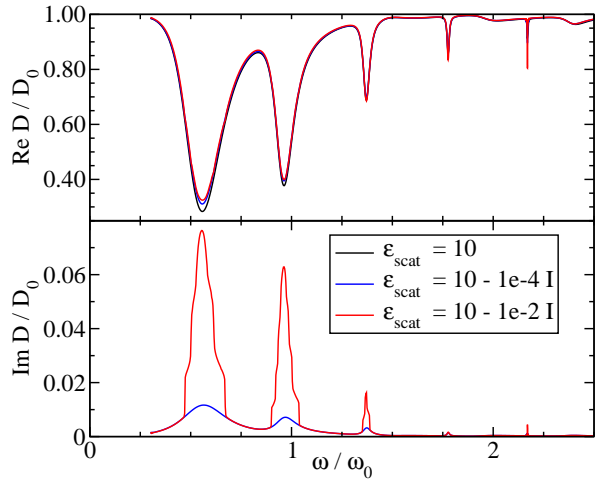} 
\end{center} 
\vspace*{-0.4cm}
\caption{
         Normalized real part (upper panel) and imaginary part (lower panel) of the diffusion coefficient $D(\Omega=0)$ 
	 for different values of  optical gain as indicated
	 as a function of the dimensionless frequency. For the displayed frequency range,
	 the gain value ist below threshold.
         }
\label{D_REAL.eps}
\label{D_IMAG.eps}
\end{figure}

In the upper panel of Fig. \ref{D_REAL.eps}, the real part of the diffusion coefficient for
a non-dissipative system (black line) is compared to systems 
exhibiting gain (colored lines). The diffusion constant has been renormalized 
to its bare coefficient, {\em cf.}  Eq. (\ref{Dbare}).
As already discussed,  small gain as compared to threshold gain
disadvantages localization, whereas with increasing gain
also the diffusion increases.
For the discussed gain values this effect is inverted within higher
resonances, because they are much closer to threshold,
where gain narrowing has already overcome this suppression.
Although the effects of gain on transport are rather small,
the small but finite gain introduces a completely new feature, 
an imaginary part of the diffusion constant at zero internal frequency,
i.e. an imaginary part to the dc diffusion coefficient or likewise
to the dc conductivity.
The normalized imaginary part of $D$ is displayed in Fig. \ref{D_IMAG.eps},
normalized to the bare diffusion coefficient $D_0$.
In the next subsection it will be shown how
this finite imaginary part  may give rise to intensity oscillation within the sample.

\begin{figure}[t] 
\begin{center} 
\vspace*{1.5em}
\includegraphics[width=0.90\linewidth]{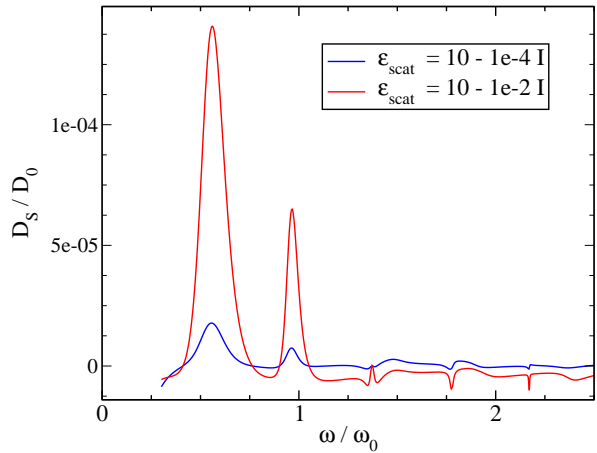} 
\end{center} 
\vspace*{-0.4cm}
\caption{
         Normalized extra contribution to the  diffusion coefficient $D_s/D_0$ for different values of the optical gain 
	 as a function of the dimensionless frequency. 
         }
\label{D_S_D_0.eps}
\end{figure}

The additional contribution $D_s$ from Eq. (\ref{D_B_NR}) to the diffusion constant 
provided by the amplifying scatterers
is presented in  Fig. \ref{D_S_D_0.eps}. This contribution is a direct consequence
of the photonic Ward identity, Eq. (\ref{Ward}), and establishes the conserved energy density.
However, for reasonable gain values considered here, this correction is seen to be orders 
of magnitude smaller than the diffusion constant, therefore its influence on the transport 
properties is only weak.

\begin{figure}[t] 
\begin{center} 
\vspace*{1.5em}
\includegraphics[width=0.90\linewidth]{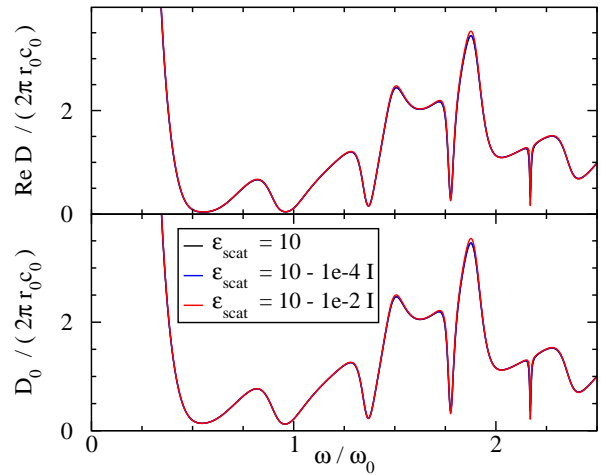} 
\end{center} 
\vspace*{-0.4cm}
\caption{
         Real part of  diffusion coefficient $D(0)$ (upper panel) and the bare diffusion coefficient $D_0$ 
	 (lower panel) for different values of the optical gain 
	 as a function of the dimensionless frequency. The diffusion coefficients are  shown in 
	 natural units ($2\pi r_0 c$), $r_0$ is the scatterers radius and c is the vacuum speed of light.
         The influence of optical gain on diffusion is seen to be small.
         }
\label{D_REAL_absolut.eps}
\label{D_REAL_0_absolut.eps}
\end{figure}

For completeness and later use we also display the bare diffusion constant  $D_0$ and the real part
of the full diffusion coefficient.  
The bare diffusion constant as shown in Fig. \ref{D_REAL_0_absolut.eps}
exhibits strong variations as function of frequency but only small variation  
with increasing optical gain. 
The full diffusion coefficient in Fig. \ref{D_REAL_absolut.eps} follows closely 
the behavior of the bare diffusion, as already indicated in Fig. \ref{D_REAL.eps}.
For very small and also for large frequencies the difference is negligible, 
visible effects are found within an intermediate frequency range only.

\subsection{Length and Time Scales}

Within disordered systems there exist different length or time scales, related to both
single and two-particle quantities.
Additionally, a geometrical mean  distance between each two scatterers $r_m$ can be defined 
by $\frac {r_m} {r_0} = \sqrt[3]{\frac{4\pi}{3\nu}}$, where $r_0$ is the scatterer's radius
and $\nu$ the filling fraction. For a filling fraction of $\nu = 30\%$ 
this ratio becomes $\frac {r_m} {r_0} = 2.41$. 

The most important single particle length is the so-called {\em scattering mean free path} $l_s$
defined in the Green's function
\begin{eqnarray}
\label{Green_ls}
G_{\vec{q}} (\omega) 
=
\frac
{1}
{\frac{\omega^2}{c^2}\epsilon_0 -q^2 - \Sigma (\omega)  }
\end{eqnarray}
where the imaginary part of the self-energy introduces the decay length $l_s$
\begin{eqnarray}
q  &=& \frac {\omega}{c}\sqrt{\epsilon_0} \longrightarrow {\rm Re } (q) + \frac {i}  {2 l_s }    \\
l_s &=&            \frac
                  {1}
                  { 2 {\rm Im } (\sqrt{q^2+i{\rm Im } \Sigma (\omega) } )     }
\end{eqnarray}
The decay length may equivalently be represented as a life time of the corresponding k-mode.

\begin{figure}[t] 
\begin{center} 
\vspace*{1.5em}
\includegraphics[width=0.900\linewidth]{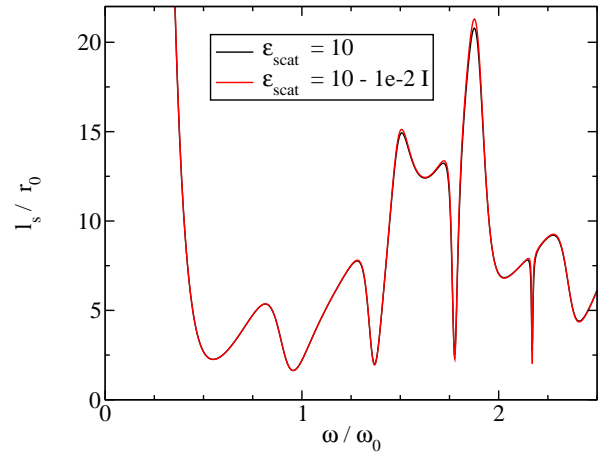} 
\end{center} 
\vspace*{-0.4cm}
\caption{
         Displayed here is the single particle scattering mean free path $l_s$
	 for  amplifying scatterers and conserving scatterers. The difference 
	 between the two is seen to be small and mainly manifests itself 
	 in narrow dips and peaks.
         }
\label{Ls.eps}
\end{figure}

In Fig. \ref{Ls.eps} the scattering mean free path is shown as a function of frequency.
The strong variation with frequency is known to be typical for the low density 
approximation \cite{Tig92,Tig93,Bus95} as used in this publication. 
The minor dependence on gain is in agreement with the 
previous subsection and mainly established in slightly more pronounced dips and 
enhanced peaks.

Before proceeding, we want to emphasize that in case of real dielectric constants the
scattering mean free path  sets the scale determining the loss
due to scattering  out of a given k-mode. Whereas in case of gain media the originally k-mode
experiences also an amplification. In this way a competition is established between 
scattering and gain. Once the optical gain is strong enough to compensate the scattering loss,
i.e. ${\rm Im }  \Sigma (\omega) = 0$, that fact is interpreted as the
crossing of the laser threshold \cite{Cao00,Cao03b,cao05,Lagendijk_mie}. This particular case  consequently defines the 
range of validity of the presented theory, which cannot describe the onset of the 
laser dynamics. The gain coefficient is therefore to be chosen such that the system remains below
its threshold gain value. This will be discussed in detail in subsection C below.

\begin{figure}[t] 
\begin{center} 
\vspace*{1.5em}
\includegraphics[width=0.900\linewidth]{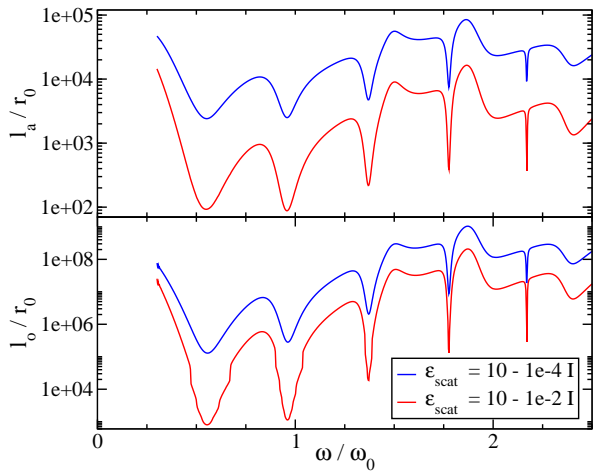} 
\end{center} 
\vspace*{-0.4cm}
\caption{
         Displayed in the upper the two particle amplification length $l_a$
	 for different values of the optical gain 
	 as a function of the dimensionless frequency.\\
	 Displayed in the lower panel is the two particle oscillation length $l_o$
	 for different values of the optical gain 
	 as a function of the dimensionless frequency.
         }
\label{La.eps}
\label{Lo.eps}
\end{figure}

Let us now  return to the discussion of the intensity and the scales related to it.
The two-particle Green's function as given in Eq. (\ref{P_E} ) contains two obvious  scales 
originating solely  from finite values of the gain coefficient. 
These length scales may be defined by
\begin{eqnarray} 
\label{def_ell_a}
\ell_a = \frac {2\pi}{{\rm Re } (\sqrt{1/\xi_a^2}  )}\\
\ell_{osc} = \frac {2\pi}{{\rm Im } (\sqrt{1/\xi_a^2}  )}
\end{eqnarray}
where $\ell_a$ represents the amplification or absorption  length of the intensity
and $\ell_{osc}$ marks the length over which the intensity oscillates,
where $\xi_a^2$ has already been defined in Eq. (\ref{xi_a} ).
The corresponding time scales may then be defined as
\begin{eqnarray} 
\label{def_tau_a}
\frac 1 {\tau_a}     &=& \frac { D} { \xi_a^2} \\
\frac 1 {\tau_{osc}} &=&       Q^2 {\rm Im } D 
\end{eqnarray}
The amplification or growth length $\ell_a$ is displayed in Fig. \ref{La.eps}
as a function of the external light frequency.
The single scatterer Mie resonances are clearly visible as well as
strong dependence on the gain value.
However, even for the strongest presented gain, the magnitude of the
amplification length remains at least an order of magnitude larger than
the corresponding scattering mean free path, shown in Fig.  \ref{Ls.eps}.
Additionally, we have plotted the oscillation length $\ell_{osc}$ in
Fig. \ref{Lo.eps}. As compared to the amplification length 
the resonant character of the scattering is 
even more pronounced and $\ell_{osc}$ of course also strongly depends on the gain value.
The magnitude is even significantly larger than the amplification length 
$\ell_a$. 
This fact may represent a large obstacle in experiments. 
The physical picture presents itself now as the following, 
if one measures the intensity distribution between
two points in coordinate space 
in the sample at a given distance $r$, i.e. 
at a finite value of $Q$ in Eq. (\ref{P_E}), one will measure a diffusive intensity decay 
as characterized  by the real part of the diffusion constant plus an exponential
amplification due to $\ell_a$ and additionally there is an  amplitude modification 
proportional to $\cos(2\pi \ell_a / \ell_{osc} )$ in case  one chooses the measuring 
distance to be  $\ell_a$. This means, in general 
even though the intensity has experienced an exponential increase by a factor of $e$,
this oscillating modulation factor 
of the intensity is close to unity and therfore possibly hard to detect.

\begin{figure}[t] 
\begin{center} 
\vspace*{1.5em}
\includegraphics[width=0.900\linewidth]{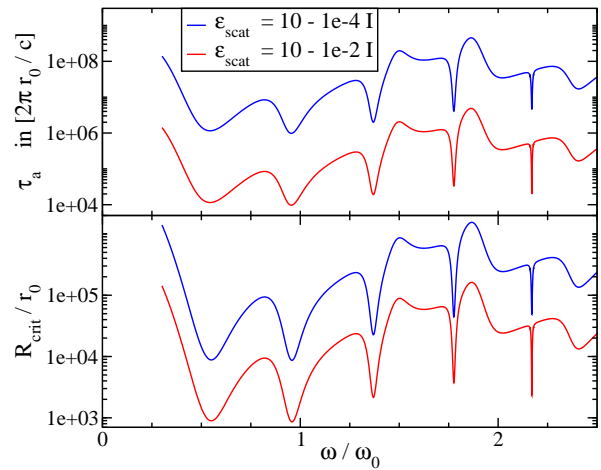} 
\end{center} 
\vspace*{-0.4cm}
\caption{
         Displayed in the upper panel is the two particle amplification time $\tau_a$
	 for different values of the optical gain 
	 as a function of the dimensionless frequency.\\
	 In the lower panel we show the critical length $R_{crit}$, characterizing
	 the threshold volume in which the intensity experiences a laser-like 
	 growth behavior.
         }
\label{Tau_a.eps}
\end{figure}

Slightly changing the point of view let us now turn to the
time scales. In the upper panel of Fig. \ref{Tau_a.eps}  we  
present the gain induced growth time $\tau_a$  defined 
in Eq. (\ref{def_tau_a}). For obvious reasons, $\tau_a$ displays a 
qualitatively similar behavior as the above discussed $\ell_a$, 
i.e., for frequencies within the scattering (Mie) resonances, the time
to exponentially increase the intensity is significantly smaller than
for frequencies outside this range. Which reflects the fact
that the gain coefficient  is confined to the scatterers volume only.

Using  the gain induced growth rate $\tau_a$ as defined in Eq. (\ref{def_tau_a}),
the intensity Green's function Eq. (\ref{P_E} ) may  now be rewritten as
\begin{eqnarray} 
\label{P_E_tau}
P(Q,\Omega) = 
\frac
{\alpha}
{-i\Omega + iQ^2{\rm Im } D + Q^2{\rm Re } D - 1/{\tau_a} }
\end{eqnarray}
where the coefficient $\alpha$ may symbolically contain all  the factors explicitly  
shown and discussed in Eq. (\ref{P_E}).
By inspection of the above equation, Eq. (\ref{P_E_tau}) and comparison
with the Green's function Eq. (\ref{Green_ls}), it is to  be recognized
that the energy density $P(Q,\Omega)$ also exhibits  a laser-like threshold behavior 
in complete analogy to the single-particle Green's function.

Before discussing this threshold behavior in detail,
we want to remind that our theory started with calculating
the electrical field-field-correlator at different positions and 
frequencies Eq. (\ref{bethe_eq}) eventually leading to the 
evaluation the two particle Green's function given in  Eq. (\ref{P_E}).
This means, the momentum $Q$ appearing in  Eq. (\ref{P_E}) represents in Fourier space this 
relative position within the sample. In three dimensions the momentum $Q$ therefore defines 
a volume unit within the sample.
This volume is carefully to be distinguished from all over length scales 
e.g. the sample volume etc.
It is merely the volume within one considers correlation effects of the 
diffusing behavior of the intensity.

In analogy to ${\rm Im } \Sigma = 0$ in the single particle Green's function
the threshold condition for the energy density now reads  as follows
\begin{eqnarray}
  Q^2{\rm Re } D - 1/{\tau_a} &\ge& 0  \\
\Leftrightarrow 
\frac
{4\pi^2}
{R_{crit}^2}
{\rm Re } D
- 1/{\tau_a} &=& 0  
\end{eqnarray}
leading to the definition of  a third, a critical, length scale
\begin{eqnarray}
\label{def_R_crit}
R_{crit}
=
2\pi \sqrt{\tau_a {\rm Re } D} . 
\end{eqnarray}
This length describes the volume over which the energy density or intensity
can compensate the diffusive loss by amplification due to the 
finite optical gain.

As in the above described case of a single particle Green's function,
the value $R_{crit}$ marks the point in parameter space where the laser threshold 
has been crossed.
From the theoretically and experimentally known behavior of a lasing 
Mie sphere \cite{Lagendijk_mie,Nussenzveig,Vahala,Lai}
 and the theoretical 
description of the self-energy $\Sigma(\omega)$ by the single scatterer t-matrix\cite{Tig92,Tig93},
and the definitions in  Eq. (\ref{def_tau_a}) and Eq. (\ref{xi_a})
it becomes clear  that in the limit of reaching the laser threshold 
within a Mie resonance $\tau_a$ is approaching zero,
$\tau_a \rightarrow  0$. And therefore the corresponding critical volume
of the light intensity becomes point-like. 
In situations where the optical gain is still below its threshold value
with respect to the single scatterer Mie resonance,
there is consequently a finite $\tau_a$ and therefore  a finite critical volume described by
$R_{crit}$. 
In the limit of large gain it becomes clear from the definitions in
Eq. (\ref{def_tau_a}) and  Eq. (\ref{def_ell_a}) 
that   $R_{crit}$ in  Eq. (\ref{def_R_crit}) approaches 
the same order of magnitude as  $\ell_a$.
Which is perfectly meaningful, since in this limit the smallest and all-dominant length scale
is set by the amplification length. For the most interesting intermediate range,
we show the critical length $R_{crit}$ in Fig. \ref{Tau_a.eps}
as a function of light frequency for different gain strengths.
Already for the  gain values discussed in this publication, which are significantly below threshold,
the ratio $R_{crit} / r_0$ becomes as small as approximately  xxx, 
{\em c.f.} lower panels of Figs. \ref{Tau_a.eps} and \ref{Functions_of_gain.eps}.

Finally, we emphasize that, once a Mie resonance is close to lasing or the gain is very strong
the growth time $\tau_a$ may become small and therefore the critical distance 
$R_{crit}$ may also become 
small as shown in the lower panel of Fig. \ref{Tau_a.eps}. 
In general the length $R_{crit}$ is  not restricted  to values above e.g. the 
single particle scattering mean free path. This is because the underlying physics is
not scattering but frequency independent amplification.

\subsection{Gain and Laser Threshold}

\begin{figure}[t] 
\begin{center} 
\vspace*{1.5em}
\includegraphics[width=0.900\linewidth]{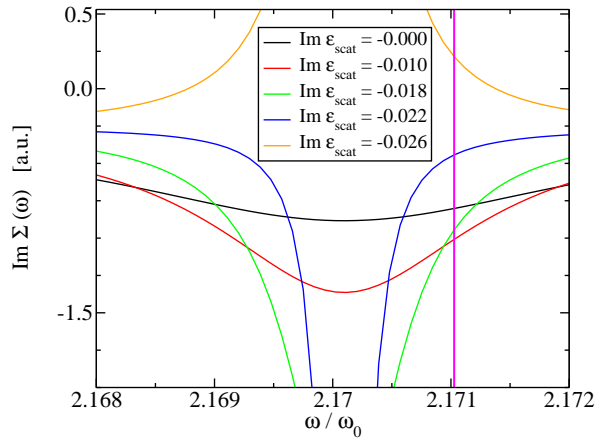} 
\end{center} 
\vspace*{-0.4cm}
\caption{
         The imaginary part of the self-energy in the vicinity of the fifth 
	 Mie resonance for different but fixed  values of gain. Increasing 
	 gain clearly yields narrowing and deepening until the laser threshold 
	 is reached, the curve approaches the shape of a delta-function.
	 Beyond this point the theory is not valid anymore due to physical reasons. 
	 For completeness we also show the last curve ${\rm Im } \epsilon_{s} = -0.0026$ 
	 representing a system which has obviously crossed the threshold.
	 The vertical line marks the frequency  discussed in Fig. \ref{Functions_of_gain.eps}, see also text below.
         }
\label{Imag_Sigma_close_Resonance.eps}
\end{figure}

\begin{figure}[t] 
\begin{center} 
\vspace*{1.5em}
\includegraphics[width=0.900\linewidth]{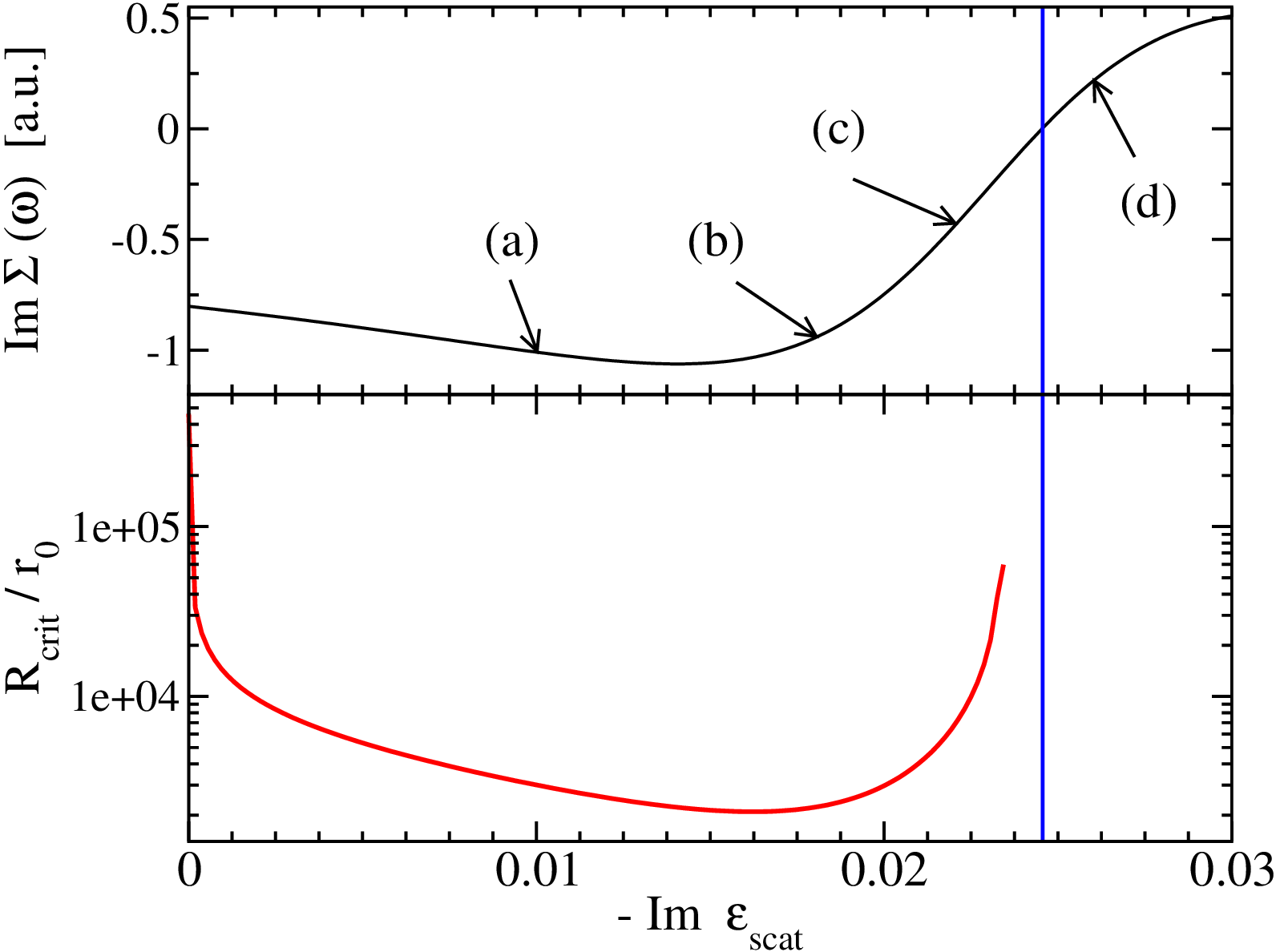} 
\end{center} 
\vspace*{-0.4cm}
\caption{
         The upper panel shows the imaginary part of the self-energy at a fixed frequency ($\omega / \omega_0 \simeq 2.1710269$)
	 as a function  of gain, {\em c.f.} Fig. \ref{Imag_Sigma_close_Resonance.eps}. 
	 The vertical line signalizes the zero of ${\rm Im } \Sigma $. 
	 From Fig. \ref{Imag_Sigma_close_Resonance.eps} and the text below, it is clear
	 that this zero indicated that the laser threshold has already been crossed.
	 The points (a) through (d) correspond to $- {\rm Im} \epsilon_{scat} = \{0.01, 0.018, 0.022, 0.026 \}$,
	 i.e. to the curves shown in Fig. \ref{Imag_Sigma_close_Resonance.eps}. The intersections with the vertical line there
	 yield the values shown here.\\
	 The lower panel displays the critical length $R_{\mbox{\tiny crit}}$ as a function of gain.
	 Due to the off-resonant frequency,  $R_{\mbox{\tiny crit}}$ has a local minimum
	 and increases again with increasing gain.
         }
\label{Functions_of_gain.eps}
\end{figure}

In this last subsection, we want to  closer discuss the appearance of a
laser threshold within our theory.
Since we approximate the single particle self-energy by the 
single particle scattering matrix calculated within Mie theory, 
we first recall some basic and well known facts. 
The resonant features representing the resonant scattering modes, arise
due to poles in the scattering coefficients, forming the t-matrix.
These poles, or zeros of the coefficients' denominators, occur at 
complex frequencies, the closer the pole happens to be to 
the real frequency axes the more pronounced is the  feature, i.e.
the resonance becomes narrower and deeper.
The effect of optical gain modeled as an imaginary part of the dielectric function,
as it is done in this paper, is to lift the poles, i.e. it shifts the complex poles
towards the real axes. In this way a gain narrowing is observed. 
The gain value corresponding to infinitesimal width of the resonance is believed to
correspond to the experimentally observable laser threshold \cite{Lagendijk_mie}. 
This situation coincides with a scattering pole right on the real frequency axes,
i.e. the scattering resonance approaches the limiting shape of Dirac's delta function with negative sign.
Using gain values above threshold causes poles in the upper complex frequency half-plane,
constituting unphysical behavior due to the neglected laser dynamics. 
If the complex frequency poles are  in the upper half-plane, the theory still predicts a resonance feature,
just with the "wrong" sign, i.e. instead of dips in the retarded self-energy one now observes peaks, see also
Fig. \ref{Imag_Sigma_close_Resonance.eps}.
Additionally the larger the gain, the less pronounced the feature becomes,
because the poles are then pushed away from the real axes in the complex frequency plane.
This contains the risk of utilizing much to high gain values, with poles in the negative complex frequency
half-plane being at large distances to the real axes and therefore yielding very weak features
that do not necessarily break causality of the single particle Green's function for instance
and might be overlooked at first glance.

To illustrate the above discussed  subject in and out of resonance, we considered a system 
slightly off-resonant but close to the fifth 
Mie resonance as depicted in Figs. \ref{Imag_Sigma_close_Resonance.eps} and \ref{Functions_of_gain.eps}.
In Fig. \ref{Imag_Sigma_close_Resonance.eps} the imaginary part of the self energy as a function 
of frequency is presented, the
quantity describing both single particle scattering and amplification, as discussed earlier.
For different gain values the narrowing and deepening of the resonance is clearly visible as well
as the difference of systems below and above threshold. The vertical (magenta) line marks
the single, off-resonant frequency for which we study the self-energy (upper panel) 
and the critical length $R_{crit}$ (lower panel)
as a function of increasing gain, as shown in Fig.  \ref{Functions_of_gain.eps}.

The behavior of ${\rm Im } \Sigma$ is easily understood 
by comparison with Fig. \ref{Imag_Sigma_close_Resonance.eps}. It is to be noted that neither
the local minimum nor the zero define the laser threshold.
The behavior of  $R_{crit}$ as defined in  Eq. (\ref{def_R_crit}) is then shown in the lower 
panel of  Fig.  \ref{Functions_of_gain.eps}. The final increase is a consequence from the gain
narrowing of the resonance. This decreases both the single particle scattering rate and
the intensity emission rate $1/\tau_a$ defined in Eq. (\ref{def_tau_a}) 
and therefore increasing the critical volume.
Following the above given line of arguments,
it becomes clear that once a frequency closer to the resonance is chosen,
the minimum value of  $R_{crit}$ decreases because the scattering rate and the intensity emission 
rate both increase. Even for a finite spectral width of the resonance, 
i.e. for a system below threshold, the value of the critical length may become quite small.

\section{Conclusion}

In conclusion, we have  presented a 
semi-analytical theory for scalar waves propagating in random, dissipating, i.e. also amplifying,  
media. The focus has been put on the influence of localization effects and finite gain on
intensity transport in general. We found that for reasonable magnitudes of gain, the impact on
transport quantities as e.g. the real part of the diffusion constant, scattering mean free path etc.
is rather small. However, the gain introduces three new length scales natural to such systems,
an amplification length $\ell_a$, an oscillation length $\ell_{osc}$ and a critical length $R_{crit}$. 
The latter describing
the critical volume in which the intensity experiences a laser-like threshold behavior.
The former two length scales constitute a growth length competing the diffusive loss of the intensity and
the oscillation period of the intensity, respectively. Due to its comparably large magnitude, this 
oscillation length might be difficult to detect in experiments.
We point out, that the critical length, or equivalently  volume, has no lower bound other than zero 
since, e.g. for a Mie resonance reaching its 
lasing threshold this volume becomes point like. For cases below threshold gain within the Mie scatterers or
off-resonant light, the critical volume is finite and strongly influenced by
the gain coefficient.

Acknowledgments - The authors acknowledge for support the Karlsruhe School of Optics \& Photonics (KSOP)
(R.F.) and the SFB 608 (A.L.). For valuable discussions they want to thank
 Johann Kroha and  Kurt Busch.

\end{document}